# An Agent-Based Model of Message Propagation in the Facebook Electronic Social Network

Hamid Reza Nasrinpour, Marcia R. Friesen, and Robert D. McLeod, *Member, IEEE*

*Abstract*—A large scale agent-based model of common Facebook users was designed to develop an understanding of the underlying mechanism of information diffusion within online social networks at a micro-level analysis. The agent-based model network structure is based on a sample from Facebook. Using an erased configuration model and the idea of common neighbours, a new correction procedure was investigated to overcome the problem of missing graph edges to construct a representative sample of the Facebook network graph. The model parameters are based on assumptions and general activity patterns (such as posting rate, time spent on Facebook etc.) taken from general data on Facebook. Using the agent-based model, the impact of post length, post score and publisher's friend count on the spread of wall posts in several scenarios was analyzed. Findings indicated that post content has the highest impact on the success of post propagation. However, amusing and absorbing but lengthy posts (e.g. a funny video) do not spread as well as short but unremarkable ones (e.g. an interesting photo). In contrast to product adoption and disease spread propagation models, the absence of a similar "epidemic" threshold in Facebook post diffusion is observed.

## I. Introduction

Marketers have widely accepted the importance of Word-Of-Mouth (WOM) for a product success [1]-[4]. For example, Philips in 2006, Hewlett-Packard (HP) in 2008, Microsoft in 2009 and Ford in 2009 all developed different types of word-of-mouth seed marketing campaigns to promote their sales [5]. The objective is to seed a marketing campaign with the intention of fostering message propagation or spread. This kind of viral marketing is not necessarily only for promoting a product; it can also help in acquiring new members or broadcasting a message or trends in general. In comparison with traditional marketing, WOM has a longer lasting impression to a new member [6]. Moreover, word-of-mouth can work well in cyberspace such as online communities, emails, product ratings, blogs, forums and electronic social networks. In particular, online WOM could be more attractive to companies because of associated lower costs, wider accessibility, immediate distribution, ease of use [7], and better tracking analytics. The importance of online social networks for venders to advertise their products can also be verified by considering the numerous blogs and publications regarding online social networks in marketing science [8]-[13].

Among online social networks, Facebook is currently the most well-known social network on the internet. However, the answer to the mysterious question of how to make a post go viral on Facebook still remains somewhat of an enigma. There are many recommendations, anecdotes and hints on different blogs to help one reach maximum influence. Apart from the complex news feed algorithm of Facebook, including a very large network of different people with complex psychologies and so many soft factors, a better understanding of message propagation mechanisms through this type of online social network is needed. Some detailed studies of the Digg social news website [14] and Twitter social network [15] exist that pay specific attention to the rules and structure of those social networks. However, many of the other works in the context of social networks and diffusion are either too generic or at a macro-level aggregate, without utilizing a detailed model of an online social network (see section IV).

In this paper, the Facebook message propagation process is examined at a micro-level. We focus on two properties of a Facebook post (post length and post interest) and one attribute of the sender (friend count within Facebook). As an instance, an attempt is made to determine which of the two properties of post length and post content (interest) plays a more impactful role in the success of post diffusion. By inspecting Facebook at the level of individuals, a better understanding of the underlying dynamics of message propagation process in Facebook is hopefully obtained. Moreover, ultimately the goal is to understand the similarities and differences between the propagation of relatively intangible entities such as messages and memes and the propagation of tangible entities such as infectious diseases within a population (see section IV).

An Agent-Based Model (ABM) of aspects of the Facebook social network was created to delineate and understand the patterns of message propagation. Agent-based modeling [16], [17] is a natural way of simulating systems where individual agents (e.g. people) play significant roles. In this bottom-up approach, the system contains a set of autonomous individuals, i.e. agents, interacting based on a set of rules within an environment. From the micro-level inter-actions of

The authors gratefully acknowledge financial support from the NSERC Discovery Grants program and the University of Manitoba for GETS funding.
H. R. Nasrinpour and R. D. McLeod are with the Department of Electrical and Computer Engineering, University of Manitoba, Winnipeg, MB R3T 5V6, Canada
(e-mail: Hamid.Nasrinpour@umanitoba.ca; Robert.McLeod@umanitoba.ca).
M. R. Friesen is with Design Engineering, University of Manitoba, University of Manitoba, Winnipeg, MB R3T 5V6, Canada
(e-mail: Marcia.Friesen@umanitoba.ca).

reading a friend's post and sharing the post among friends, the macro-level patterns of diffusion of the post evolves. This makes ABMs a suitable method of analyzing Facebook where both micro-level and macro-level analyses are of interest. In addition, we have a heterogeneous population of Facebook users, which is inherently suitable to an agent-based model where each user can have their own profile, in this case, preferences of when to sign in or share a post. ABMs are generally well suited to model social networks when either the agents or the topology of the interactions is heterogeneous or complex [18].

The idea of deploying an ABM is not only for analysing the impact of post length versus post score, but rather for making a tool potentially capable of including a variety of features to set up different experiments that empirical work cannot address. Cellular Automata as a limited form of ABM has previously been applied for modeling a seeding program [5]. In our study, similar to cellular automata, agents have an internal state machine but are not modeled as (limited) cells. The agents are people modeled as nodes within a graph. Each has its own internal parameters such as the number of friends within the social network. They can asynchronously and independently act (see section II). As an example, agents can decide when to log in/out, read and share a note among friends. The detailed behaviour of agents is explained in section II. ABMs are also inherently extendable to almost infinite levels of details, and in this case the ABM can be extended to study other possible actions by Facebook users (see section V). Depending on the scope of the research, one can add or remove rules and states to or from agents. An advantage of ABM over Differential Equation (DE) or Statistical models is its lack of complex math, which means we do not have to understand relatively complex model formulations [18]. In other words, we just need to be able to describe the system and agent behaviour in detail with a set of "what-if" rules, in a problem-specific and natural lexicon. These characteristics make ABMs well suited to problems that are computationally irreducible [19]. This has however left ABMs open to critique as being more difficult to validate.

Obviously, there are also other limitations within ABMs. ABMs can easily be slow and computationally intensive. In fact, the speed performance was one of the main obstacles encountered here in obtaining the results which are presented in section III. These limitations are acknowledged in section V.

The remainder of this paper addresses the following:

- A realistic agent-based model of Facebook posts diffusion was created. This framework can easily be extended to include more features of Facebook and its users.
- In an initial set of simulations, the relative importance of each input factor such as post score, post length and publisher's friend count is compared.
- A second set of simulations explores the impact of the details of post score versus post length particularly for shorter posts like URLs or photos more specifically.
- A third set of simulations sheds light on two seeding strategies relative to a mass of users versus a few hub users.
- Generally, post content has the highest impact for information propagation within the electronic social network; however, among the posts which spread fairly well through the network, post length is of more importance than the post content (interest) and the initial seeder's friend count.
- Surprisingly, it is shown that there is no tipping point [20] for post diffusion analogous to the transition to epidemic spread observed in infectious diseases. On average, the moment a post is submitted is when it reaches its peak of the probability of being shared or read by friends.
- It is also shown, unlike product adoption or disease spread, that it is unlikely for a Facebook post to go viral and reach a fair percentage of the entire network. In this case (like other celebrity phenomena), the fact that some posts obviously do go viral may skew a typical Facebook user's perception of the probability of their own post doing so.

## II. ABM ARCHITECTURE

The ABM is implemented in the Java-Based educational version of the Anylogic software toolkit, which supports Agent-Based, Discrete Event and System Dynamics Modeling. In this section, the agent-based model, agents' properties, the structure of their environment, and the governing rules are explained in detail.

### A. The Big Picture

People using Facebook can either visit the webpage on their browser or use the Facebook application on their mobile device. In either case, once you open your Facebook profile, you may receive a list of notifications of what has previously happened since your last login. The difference is that in the second case, you can stay signed into your Facebook profile with your Facebook phone or tablet application, which results in receiving notifications when they occur. Once a person truly decides to check their Facebook profile, they usually go through the notifications and then generally switch to their news feed (home) page to see the activities and posts from friends or other Facebook pages/groups to which they are affiliated. At any time during the visit on Facebook, a user might decide to post a text note or upload a photo/video on their Facebook (wall/timeline) page. They might also copy a post previously shared by a friend and paste it on their own page in order to share it with their own friends. Through this feature, a post would spread over the network. The other common way to interact with a post is to "like" which is invoked by clicking a like button below a post. Currently, there are also many other features available on Facebook such as private messaging and applications that all act like incoming stimuli to a Facebook user to draw their attention. The current scope of this study is general posts by users on their own wall page. We recognize that this is a simplification of actual social networking via Facebook. The simplification was necessary in the first attempt at creating a model. Detailed simplifying assumptions are presented in section II.D where the rules of behaviors by agents are explained.

Herein, akin to reality, time passes continuously in minutes and seconds. Agents (Facebook users) are connected to one another within a virtual social network. Each agent, independently from all the other events, decides when to log in and when to log off. During the interval they are logged into the system, they go and check their friends' Facebook wall pages, each post one by one, until they decide to switch to another friend's page. Each post has some interest score and requires its own unique time to be read. Once an agent finds out something interesting on a friend's page, they might decide to share it again on their own page. Also at any time when an agent is online, they can publish a new post of their own. This agent is denoted the initial seeder/publisher of the post. Agents - when they are online - are able to receive notification of recent activities from their immediate friends. In the current agent-based model, this activity only includes the case where a friend shares a post on their own wall page. In this agent-based model, similar to the real Facebook where users check their notifications, upon receiving a notification by an online agent, they go through the notification and read the post shared by their friend.

*B. Agents and Parameters*

The agent-based model consists of only one type of agent which is a Facebook user or individual. Agents can create and publish different posts with two important properties of *Post Length* and *Post Score*. Each agent has a set of internal parameters including *Activity*, *Average Login Time*, *Average Post Rate*, *Friend Time* and *Friend Count*.
Wherever possible, published reports on Facebook user statistics were used to set agent parameters, and where a given parameter was reported in several references, reasonable inferences and consolidations were made [21]-[28], etc.

The Activity parameter is the main parameter that characterizes the heterogeneity in individual preferences for posting/sharing notes. In other words, each user posts new notes at a given rate, which depends on the user's Activity parameter. Also, the chance of sharing a post already shared by a friend is related to this parameter. This parameter is assigned a uniform random value from 0% to 100%. Obviously, the higher the value, the more posts the user generates. Generally, the more active users (users who post more often) are also online more often. The exact association between the frequency of activity and online time is described below.

The Average Login Time parameter indicates how many minutes a user is online in a day, on average. This parameter has a normal distribution and is used to calculate the Login Time parameter by the following formula: *Login Time* = 2 × *Activity* × *Average Login Time*. This means that a typical user, whose activity parameter is 50%, has a Login Time parameter value around the value of Average Login Time parameter. Also the formula states that active users spend more time online on average. Furthermore, the login time follows a long tail distribution that also accounts for 'lurkers' (high online time but low Activity). The exact mathematical distribution of a product of a uniform random variable and a normal random variable can be found in the Appendix.

The Average Login Time parameter has a normal distribution. There are different choices for its normal distribution properties. In 2011, Facebook Press Room reported the average Facebook user spends more than 11 hours per month on Facebook [22]. Assuming 30-day months, this means over 22 minutes per day. In addition, [23] claims the "average user spends an average 15 hours and 33 minutes on Facebook per month," which equates to roughly 31 minutes per day. There are also some reports on mobile usage, such as an average of 441 minutes per visitor in each month (i.e., 14.7 minutes per day) reported in [24]. Consolidating these sources, the average online time was set at 23 minutes per day, which may change in the future work. Considering the fact that most obsessed Facebook users spend daily average of 8 hours on the site [25], the average online time is bounded by a maximum of 10 hours per day.

Assuming a day is 1440 minutes, a login time of 23 minutes results in having a logout time of 1440 − 23 = 1417 minutes. This means if the login time for a user is calculated to be 23 minutes, they will be online for 23 minutes and will be offline for 1417 minutes in a day. However, since most people tend to check their Facebook profile more than once a day (e.g., It is reported that "on average, [a person] visits the Facebook app/site 13.8 times during the day, for two minutes and 22 seconds each time" [26]) these 23 and 1417 minutes have to be divided into different intervals. These intervals are drawn from an exponential distribution with the mean value of 1440/1417 and 1440/23 minutes per a day (1440 minutes), for the login and logout times respectively. The choice of exponential distributions for login/logout rates where the probability to login is the highest immediately after logging out may sound irrelevant. However, similar to working out at gym, people are likely to check their social profiles within a certain period of day (e.g., it is reported that "peak Facebook time is during the evening, just before bed" [26]). Although for some smartphone users with Facebook application installed on their phone, this period might be from 6 am to 11 pm. In either case, it would be safe to assume that people would not check their Facebook profiles when they are asleep! In the agent-based model, the login intervals should not be too far away from one another, and there has to be a limit to control when people log into their profile. The choice of exponential distributions attempts to keep the online intervals close to each other. In addition, a login rate of an average of 23 minutes/day compared to a logout rate of an average of 1417 minutes/day is low enough to span the daytime. The choice of the exponential distribution is also related to performance issues. Since it is the default distribution for rate triggers in Anylogic, deploying another distribution would significantly have decreased the speed of simulation. Part of the reason may be that the triggers scheduler in Anylogic can be set ahead in outer code loops of the program.

The Average Post Rate parameter defines the average number of new notes published per month by each user. It has a truncated normal distribution based on the parameters shown in Table I. The actual distribution of the number of posts is calculated as: *Post Rate = 2 × Activity × Average Post Rate*. This distribution is plotted in Fig. 1. The mathematical formula of this probability density function is derived in the Appendix. Assuming a month is 30 days, the exact time when an agent publishes a new note is drawn from an exponential distribution every time the user logs in with a rate of 30/*Post Rate*. There are several different reports on the rate of different posts in Facebook. Additionally, these numbers are intuitively known to keep changing. However, we had to adopt one of these reports, which is that in every 20 minutes, over 1 million links are shared, 1.8 million statuses are updated and 2.7 million photos are uploaded [27]. This results in having 11.9 billion posts per month. Add to that, the knowledge that Facebook had 845 million monthly active users as of December 31, 2011 [28], implies that there are approximately 14 posts per month for each person. This justifies the choice of an average of 13.8 posts per month for each user.

Each user is connected to some other users known as friends. The number of friends a user has is controlled by the Friend Count parameter. The number of seconds dedicated for each friend to check their posts is set by the Friend Time parameter. Every time a user begins checking their friend's posts, this timeout value is generated based on the associated normal distribution. Once the time is up, the user leaves the current friend and looks at another friend's recent posts. Complementary to this parameter, the Post Length parameter dictates how many seconds are required to completely read a certain post. When a user publishes a new note, this value and the Post Score parameter are calculated and assigned to the post. The Post Score parameter represents how interesting and appealing a post is.

Table I summarizes the model parameters and their distribution properties. As shown in the table, the Activity and Post Score parameter are drawn from standard uniform distributions between 0 and 1. For the other parameters, sampling from truncated normal distributions in Anylogic is employed. This kind of distribution is essentially the standard normal distribution which is stretched by the *Mean* coefficient, then shifted to the right by *Std. Dev.*, after that it is truncated to fit in [*Min, Max*] interval. Truncation is performed by discarding every sample outside this interval and taking a subsequent try.

*C. Environment*

Akin to online social networks, the environment is a graph where each node represents a user/agent whose social friends are neighbour nodes in the graph. In this subsection, we describe this graph and its properties in detail.

The Facebook network graph can be viewed as a small-world network [29], [30] as most nodes can be reached from every other by a small number of hops. Generally, however, scale-free networks are a better choice to model a social network graph, as they have a more realistic degree for the power law distribution. In fact, it is shown that scale-free networks themselves are ultra-small worlds, where the shortest paths become even smaller [31]. Yet the strict power-law distribution is not accurate enough to represent Facebook's degree distribution [32]. More precisely, if the power-law distribution of $P(k) = k^{-\alpha}$ is accepted to be the degree distribution of nodes, two sections for $1 \leq k < 300$ and $300 \leq k \leq 5000$ can be approximated by a power law with exponents $\alpha_{k<300} = 1.32$ and $\alpha_{k \geq 300} = 3.38$, respectively [21]. Therefore it was decided to synthesize the graph by directly sampling from the Facebook graph. Ideally one should generate multiple sample graphs and run the experiments on all the sampled graphs to ensure the results are not specific to just one graph. But due to processing time-limitations, only a few graphs were sampled; as all the sampled graphs shared comparable properties, one representative graph was selected for the basis of the agent-based model.

A dataset of the Facebook social graph released by the Networking Group of the University of California, Irvine [21] was used. Two datasets of uniform sampling and Metropolis-Hastings Random Walks (MHRW) were available. The Metropolis-Hastings algorithm is a Markov Chain Monte Carlo (MCMC) method to simulate a complex distribution from which direct sampling is difficult. The MHRW option was chosen, as the Facebook IDs within this dataset are consecutive numbers, which makes it easier to construct the graph. Gjoka *et al.* obtained this dataset by 28 Facebook-wide independent MHRWs in April of 2009 [21]. The dataset contains the number of friends and their Facebook IDs for approximately 957K unique users. It was not possible to

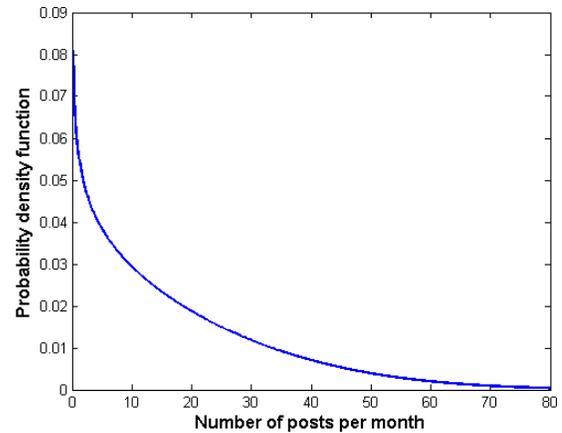

Fig. 1. Post Rate density function

TABLE I
SIMULATION PARAMETERS AND THEIR DISTRIBUTIONS

| Parameter | Distribution | Values | | | |
|---|---|---|---|---|---|
| | | Mean | Std. Dev. | Min | Max |
| Activity | Uniform | 0.5 | $1/\sqrt{12}$ | 0 | 1 |
| Average Login Time* (minutes/day) | Normal | 23 | 120 | 0 | 1080 |
| Average Post Rate* (times/month) | Normal | 13.8 | 13.8 | 0 | 300 |
| Friend Time (seconds) | Normal | 30 | 30 | 10 | 600 |
| Post Length (seconds) | Normal | 30 | 30 | 10 | 600 |
| Post Score | Uniform | 0.5 | $1/\sqrt{12}$ | 0 | 1 |

* Multiplied by 2 times *Activity* for calculating the associated rates

directly build our network structure on the dataset itself, as it was missing a large number of edges. Strictly speaking, the dataset has a uniform random sample of users in Facebook. Here, however, a dataset was required which would densely cover only some small region of the Facebook graph. Such a dataset could be obtained by a Breadth First Search (BFS) crawling method. Hence the following approach was taken:

1) The first 500 sampled Facebook user IDs were assigned to the 500 primary agents in the model.
2) Each primary agent was connected to a number of new secondary agents according to the number of friends of their corresponding Facebook user in the dataset. This resulted in a network of total 89,977 agents, but the number of edges was not sufficient.
3) Extra links between the secondary agents were inserted based on a custom distribution of all the sampled users in the dataset. This resulted in a network with a total of 7,528,164 edges.

The network was created and used (saved) for further experiments. After the second step, each node on average had only one connection, which is not at all the case in the real-world Facebook graph. The reason is that the dataset tells us the node $x$ is connected to $x_1, x_2, ..., x_n$; but hardly ever is any information available for each of $x_i$'s. As a result, the third step is necessary. The first and second steps are quite straightforward. The challenge is in the third, where creating an undirected graph of n nodes according to a certain degree distribution (here denoted distribution F) is desired. In this case n = 89,477 and the probability distribution F is given by the dataset.

Furthermore, it was desirable to create the graph in such a way that nodes with more mutual friends have higher probability of making friends than that of total random nodes. Therefore, the approach is based on the idea of common neighbours as well as the Erased Configuration Model [33]. It is defined as follows. For each node $n_i$, draw a degree $d_i$ according to the probability distribution F. Make $d_i$ half-links (or stubs) and connect them to the node $n_i$. When all the stubs are created, start the following loop procedure through the stub pool:

1) Choose three distinct random stubs $d_1$, $d_2$ and $d_3$ and remove them from the stub pool. Notice that each stub is already connected to a node.
2) Among these three stubs, find the two stubs $d_i$ and $d_j$ with more mutual neighbours between their corresponding connected nodes in such a way that they neither make multiple edges between the nodes $n_i$ and $n_j$ nor make a loop on the node $n_i$ or $n_j$.
3) If the step 2) was successful, join the two stubs $d_i$ and $d_j$ to make a connection between the two different nodes of $n_i$ and $n_j$; otherwise return one of the stubs (e.g. $d_1$) to the stub pool.

The above procedure removes two stubs from the stub pool at each repetition. Therefore, after exactly $\lfloor \sum_{i=1}^{i=n} d_i / 2 \rfloor$ iterations it terminates; it could be that one stub is left behind in the stub pool. This is quite natural in the Erased Configuration Model, and the proof of its convergence to the desired degree distribution can be found in [33].

To ensure that the graph represents the 'six degrees of separation' phenomenon, the diameter of the synthesized graph (i.e., the longest shortest path) was estimated. To measure such a statistic, 6K nodes (out of roughly 90K nodes) were randomly selected and the Breadth-First Search (BFS) algorithm was run for each node to count the number of reachable nodes at each hop. It was found that all the nodes after three to five hops reach 100% connectivity to all the rest. The percentage of reachable pairs within a certain distance is shown in Fig. 2, which is similar to the graph of degrees of separation of Facebook reported in [32]. The average distance was calculated to be 4.0074, which is comparable to 4.7 and 4.3 of the global and U.S. population of Facebook users in May 2011, respectively [32], [34].

The other statistics regarding the constructed graph, the dataset [21] and the Facebook graph [32] are illustrated in Table II, demonstrating that the dataset with a smaller number of nodes was sampled successfully. As the constructed graph has a much smaller number of nodes (but retains the same number for maximum possible number of friends), it is much less sparse than the real Facebook graph. This is part of the reason the mean distance has been reduced, compared to Facebook. Most conservatively, the synthesized graph can be considered as an acceptable example of Facebook connectivity structures of a sub-region of Facebook network graph. Nevertheless, 90K is almost the maximum possible number of users to handle during the simulations. The reason for this is that all the nodes have their own autonomous behaviour and processes, and their interaction with each other is a function of the number of edges in the graph. The time complexity of only constructing the network graph itself is O(|E|), where |E| is on the order of a million.

Finally, the complementary cumulative degree distribution function (CCDF) of the dataset and our graph is illustrated in Fig. 3, displayed on a log-log scale. As can be seen in this line graph, the distributions do not strictly follow power-law distributions, which are straight lines on a log-log plot.

*D. Rules*

In delineating the scope of the model, the following assumptions have to be made:

1) As we know, there are many different pieces of content, such as photos, videos, links, status updates, event invites, notes and etc. [35] that can be published on a Facebook wall/timeline page. However, throughout the rest of this paper the word "post" or "note" is used to indicate generic content that is posted on a wall page. This is because all content shared on a Facebook page act as incoming stimuli to a Facebook user. The objective in this study does not aim to cover the detailed properties of a successful or uneventful post. Regardless of the specific type of note, each post requires some time to be noticed by user agents. This is why the post length is of interest to us. In addition, each post is trying to convey a message to the viewer/reader. The message could be a warning, funny picture, personal news, an inspirational quote, amazing fact, etc. Each might be appealing to specific types of people. However, to keep the model simple, we considered a single scalar value to represent the general interest in a post as the post score. Each user, depending on their activity level, may share an interesting post.

2) Users cannot send private messages to one another. They can neither make comments on any post nor share any note on somebody else's wall except for their own wall; in other words, they are only able to share something on their own wall. Adding the private messaging property could have only increased the complexity of the model as the model is limited to public means of sharing a post over the network. Posting notes directly on someone else's page is not at all as common as sharing posts on personal wall pages. Incorporating these features would have demanded obtaining more statistical data about the activity behaviour of people on Facebook. Here an effort was made to not increase the number of model parameters, as tuning these parameters is one of the most challenging parts of designing an agent-based model. Even at the current model, values of some parameters are based on intuitive rational assumptions rather than actual data. Commenting on posts is a very important feature of Facebook social network. Through comments or lack thereof, a post can stay alive or die. Comments and Likes in Facebook have a direct relationship with the news feed algorithm of Facebook. In simple words, users are more likely to receive more (recent) posts on their home page from those friends to whom they had most interaction in the past. One way of making interaction with friends is via liking or commenting on their posts. The exact Facebook algorithm to rank the news feed page is unknown. In our model, we let each user agent choose which friends to have interaction with. The commenting feature is removed from the model to decrease its complexity. Adding these features back into the model could be a very nice extension of current model.

3) The only possible relationship between users is bidirectional friendship. This means subscriptions to Pages or Groups held in common between two or more agents, and any other similar features are ignored in the model. Page and groups can be considered as normal user agents in the graph with higher number of friends and activity level. In social science, they are referred to as hubs. So there is no need to distinguish pages from people in our model. Subscription to a Facebook user is similar to a unidirectional friendship. This feature was added later to Facebook. One can think of the current agent-based model as a model of Facebook in its first years without this feature.

4) Similar to the notification feature in Facebook, when a user is online, if a friend of theirs shares a note, the user will receive a notification message within the model. As a result, considering assumption 2) above, there is no way to receive or potentially save a notification for an offline user. In this model, when users go online they begin to check recent posts by friends; as such, adding a notification feature to the model would not have made a significant change to the results; but it would make a difference in time complexity of the model.

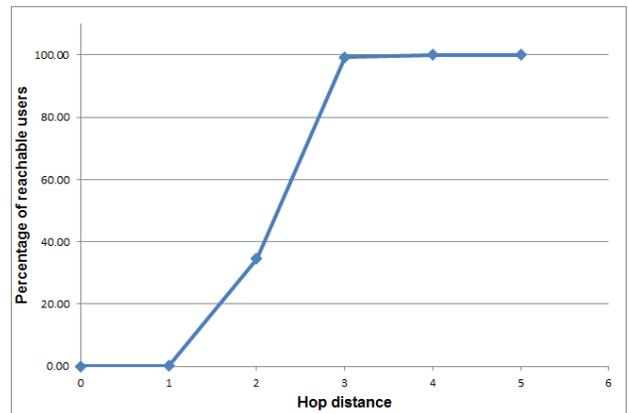

Fig. 2. Degrees of Separation: Percentage of user pairs within x hops of each other

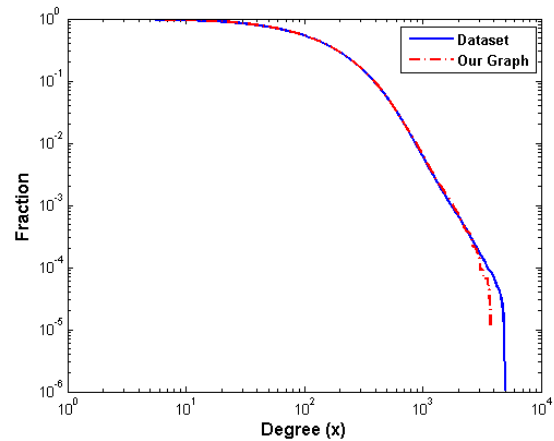

Fig. 3. Degree Distribution: The fraction of users who have degree x or greater

TABLE II
NETWORK GRAPH STATISTICAL PROPERTIES

|  | Facebook [32] | Dataset [21] | Our Graph |
| --- | --- | --- | --- |
| Mean Degree | 190 | 168 | 169 |
| Median Degree | 99 | 110 | 111 |
| Min Degree | 0 | 0 | 1 |
| Max Degree | 5000 | 4979 | 3734 |
| No. of Nodes | 721 M | 957 K | 90 K |
| Mean Distance | 4.74 | N/A | 4.0072 |

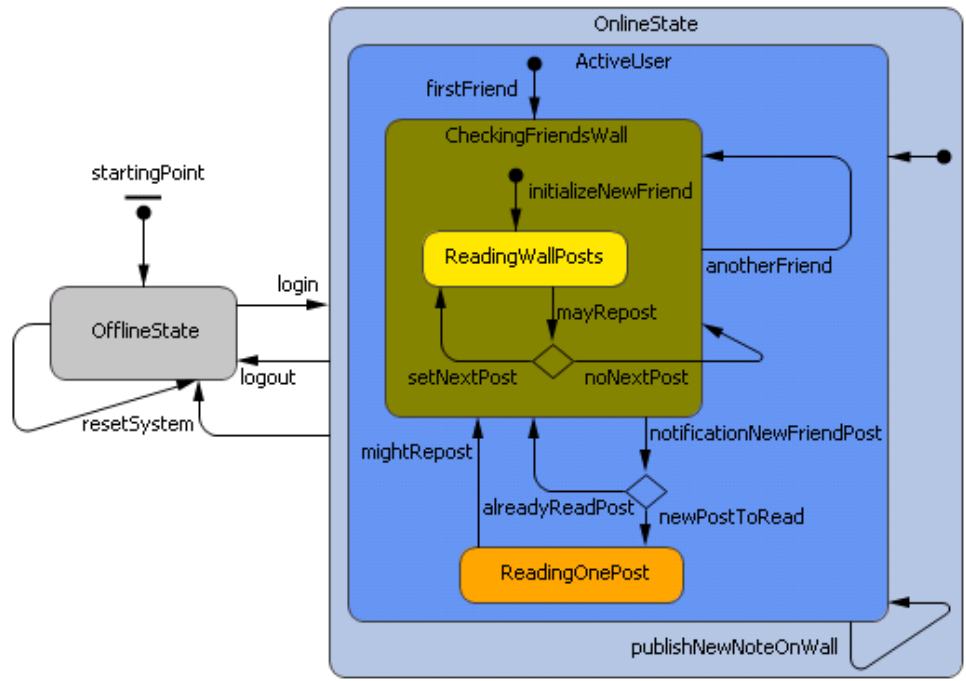

Fig. 4. User state machine

5) The network is static. This means users cannot make or remove any friendship connections so as not to change the reality-based network graph. The scope of the current agent-based model does not include effects of dynamics of the network graph. There is a viral marketing study where the evolution of network graph and changes in preferences of users for different subjects have been analyzed [36]. This heterogeneity of preferences in different topics is controlled by the activity parameter in our agent-based model. Changing a user's friends does not significantly change chances of repost. Because the activity parameter is uniformly distributed and there is no similarity between friends' activities. One might say more active people are more likely to be friends with one another. But there are many more important demographics characteristics between friends such as nationality and age which trends to be similar. Thus, the activity parameter cannot be considered as a significant factor in this list.

The agent-based model keeps an inner state for each user, controlling their behaviour. This stochastic hierarchical state machine is shown in Fig. 4. All the users are in one of the two general states of *Online* or *Offline*. Every time the model restarts, the login and logout rate for switching between these two general states are assigned to each user. The intervals when a user is in the *Online/Offline* state are controlled by login and logout rate. When the login/logout transition is triggered, they log into/log out of the system. Login transition times are drawn by an exponential distribution with the mean value of $\frac{Login\ Time}{1440}$ minutes per a day (1440 minutes). This was explained in more detail in section AII.A.

As seen in Fig. 4, all the interactions and events for a user happen when they are in the O*nline* state. Users in the *Online* state can either produce a note and share it or repost a note which was shared before by another user (which then appears on their own wall, in keeping with assumption 2) above). As mentioned earlier, each user posts new notes with a given rate based on the user's activity.

During these intervals between publishing a new note, users check their friends' wall pages. It was possible to employ a stack to sort recently posted notes for each user. However, in order to increase the level of autonomy of users, the freedom to choose which friend to see their posts is given to users. In the preliminary builds of the model, users sorted their friends based on nationality and differences in their ages. As we did not aim to analyze clustering effects of nationality and ages, in the final version, they randomly select friends to review their most-recent wall posts to speed up the simulation. This is also accordant with the Facebook policy to make a ranked list of friends' posts for each user in their homepage (News Feed). This feed is an algorithmically ranked list of friends' posts based on a number of optimization criteria [37]. In its simplest case, the feed contains recent posts of those friends with higher probability of having interaction with the user. For each friend, a normally distributed random value is given as the maximum time threshold to check their wall page. When the time is passed, the user selects next friend to check their wall posts. This loop continues until one of high-level timeouts, such as logging off or publishing a new post, happens.

As shown in Fig. 4, when a user is checking friends' wall posts, if one of their friends publishes or reposts a note, they will be notified. Consequently, the user stops their current task of checking friends' page and begins reading the new shared note, which may be considered a result of human curiosity.

Users are able to recognize a duplicate post and skip it in one second. Otherwise they would spend time in the amount of the post's length to read it.

When a user completely reads a note, they may decide to copy the note from the friend's wall page and repost it. The likelihood of this decision increases with the product of the post score and the user's activity. As an instance, a super active user with the activity value equal to 100% would definitely share an interesting post with the score of one (with probability $1 \times 100\% = 1$). One could have picked other formulas to define the probability of sharing. Multiplication, however, was found to be the simplest (and fastest) to proceed with the simulation. By this mechanism, posts propagate over the network, which is the subject of this study. It was also possible to let users share the post before they themselves read it completely. This scenario usually happens when people see a public warning that they think it might be of use to others. Again, adding a new parameter and possibility into the system would have decreased the speed of simulation. It was decided to keep the model as simple as possible in the current version.

One might ask if it is not more likely that the marginal score of a post also increases as the post is shared over and over by users. This is the case for product adoption models, mostly inspired by the classical Bass diffusion model [38] or disease epidemic modeling approaches such as the Susceptible-Infected-Removed (SIR) models [39]. In the product adoption models, it makes sense that as the number of people/friends using a certain product grows, others would become more interested in purchasing the product as well. Also, in epidemic models, as the number of infected people with a certain virus increases, the chance of transmission of the virus to others increases. However, propagation of a meme through a social network might be different. Similar to product adoption, some individuals might find sharing a popular post very cool or helpful, such as posts asking to unify people on some belief or giving alerts. On the other hand, some may consider repeated exposure to a specific peace of content as being boring or démodé. The second case is verified in Digg social news website where people have a tendency of not sharing repeated news [40]. Thus, as herein all various types of content are modeled as posts with different scores, we neither increase nor decrease post score as it disseminates through the users; each user, depending on their activity level, might pick a post and share it. A post score, which is fixed, can be determined by external factors which are not part of this study. For example, assuming a post is a product advert, a combination of the market conditions and psychological and sociological characteristics of consumers would determine this factor for each person. There are different studies concerning this aspect such as the decoy effect in [41]. Having implemented an agent-based model, they showed how an individual consumer's judgment on purchasing a product changes from brand A toward brand B after the introduction of a decoy brand, in a competitive market.

In the current simulations, for a given post, the post length time must be passed until we can conclude that the user has read the post. For instance, in the situation where a user decides to log out while reading a post, the post will not be checked as a read post.

### E. Verification and Validation

As per Rand and Rust [42], verification of a model ensures the simulated model matches the conceptual model. This procedure is mainly through documentation, program testing and test cases [42]. In the present case, the exact written assumptions and rules of behaviour defined above were coded. The programming code is also extensively commented. Each function and module of the agent-based model was solely tested on a small network of users to receive a known output for a given input. After all the debugging was done, corner cases with extreme values (such as full connectivity, no connectivity, zero activity-level users, one activity-level user, zero post score, one post score, etc.) were run.

As per Rand and Rust [42], the four major steps of ABM validation were followed. These are micro-face validation, macro-face validation, empirical input validation and empirical output validation. At the micro-face level, actions of users are a limited form of a real Facebook user's possible actions. Also, the mechanism through which a post propagates is a type of cascade model corresponding to the real world. This paradigm has been validated by various studies over the past. At the macro-face level, it is observed that most posts will not be shared by a friend, which is the same scenario for a typical Facebook user. Also, our aggregate pattern of the post share distribution on-face is similar to Facebook and other social network statistics. At the empirical input level, the ranges of all the parameters are drawn from either Facebook network statistics or reasonable assumptions. Further explanation of the input parameters is discussed in section V.

Relative to empirical output validation, it is not possible to exactly validate the agent-based model against reality at this stage, nor is this the claim in this work. The correct way is to run our experiments on Facebook; firstly however, it is not clear how to set a post's probability of being shared (i.e. its score), therefore it cannot be validated in this way in practice, apart from the fact we should be able to monitor exactly how many users have indeed read a post completely. Again this cannot be measured in practice. We may be able to only distinguish if someone has seen the post for at least a few seconds. Unless Facebook or similar social media service providers were using, for example, built-in counters to calculate how many seconds each user spends on a specific post, which is very unlikely for non-video posts; and if there exists such a counter, it would only work when you actually open a post (e.g. a photo) but could not be calculated when you are at your Facebook home page displaying more than one post, as it is not known which post you are exactly looking at. Having said that, it is still possible to demonstrate that real world data are possible outputs of our agent-based model, meaning that our average results match average results in reality.

A set of experiments was performed in order to validate the results here against the Facebook statistics reported in [37]. According to the statistics for users' most recent post, "the median post reached 24% of a user's friends (mean = 24%, SD = 10%)," provided that the most recent post was at least 48 hours old. Their population size is 589 different users with the median friend count of 335 (mean = 457, SD = 465).

9To do such an experiment, 20 unique users were selected such that their friend counts were drawn from a normal distribution with the same mean and standard deviation. As a result, the population of our selected users had a median friend count of 335 (mean = 464, SD = 337) which is quite close to the sample they used in [37].

Then their Activity parameter was set to be 1%, to prevent them from publishing any other note, since we wanted to perform the test for their most recent post; after that a post with score of zero and a short length of 10 seconds was published by the user. The score is set to zero to make sure the post does not spread over the network as we are interested in the number of immediate friends who read the post. The length was set to 10 seconds as this is our minimum post length at the current model. As mentioned above, the statistics reported by Facebook is in fact the number of people who have seen the post, although not necessarily read a long post completely. Then we ran the model. The model ran for 24 simulated hours as an initial warm up stage. Then a chosen user published a specific post. The simulation ran for another 48 simulated hours, after that the result was saved. The whole experiment for all 20 users was repeated two times, for a total of 40 runs. The results were very close in both runs for each user. Real time computation for each run was 9-12 hours on workstations equipped with an Intel Xenon CPU W3679 @3.2GHz with 16 GB RAM or higher configuration.

We observed that the median perceived audience size was 19% of a user's friends (mean = 25%, SD= 18%), which is near although slightly less than the expected 24% median of Facebook [37]. One difference in our experiment and Facebook statistics is that our results are the statistics after exactly 48 hours, whereas the Facebook ones are the statistics after at least 48 hours. So it is reasonable to reach a lower percent of immediate friends. Strictly speaking, in our agent-based model, the post life time is defined to be the last moment when the post is read. The average and median post life time in this set of experiments was 47.08 and 46.81 hours, respectively.

## III. SIMULATION STUDIES

In this section, the agent-based model was run with different input settings to explore the impacts of each factor in the post diffusion process.

In the current model, the targeted input parameters include: (1) Post length; (2) Post score; (3) Friend count which is the number of friends of the first publisher of the note. As each minute in the simulated world takes 7-10 seconds to run computationally in the real world, we have limited our initial exploration to these three parameters. Future simulations will explore the impact of other parameters, including the day/time to disseminate the note over the graph.

The following statistics as the model output were recorded: (1) The number of users who have read the note; (2) The number of users who have reposted the note; (3) The times when the note was read / reposted; (4) The last time when the note was read as its life time.

The Analysis Of Variance (ANOVA) procedure between different scenarios is employed to find the importance of each parameter relative to the spread of a message within the online social network. In some cases, multivariate regression was run to test the magnitude of each factor. However as the outputs of an agent-based simulation should not be interpreted quantitatively [17], [18], the numerical values of a linear regression coefficients are interpreted qualitatively. This means even if we obtained numbers as the magnitude of importance of each parameter, they have to be discussed at the qualitative level.

### A. Study 1: General Insight on Input Parameters

Within the first study, a total of 19 unique scenarios (simulations) were set up and each simulation was repeated 20 times, for a total of 380 runs for eight simulated days each. The real-world computing time required for each run was one to two days. The total computing time required for these simulations was over 380 days. In each scenario, our chosen user publishes a certain post after one simulated day of warmup phase, then one week after the spread, the outputs are saved and the simulation ends.

The results of the first set of simulations are illustrated in Table III. According to the first (six) rows of the table, it is immediately clear that as long as the number of repost is near zero, the post score does not have much impact on the number read. Because the post is not shared by anyone except for the first publisher, the number read is directly related to the number of immediate friends of the publisher, which is trivial. For the simulation IDs 1-6 plus IDs 9-11, with the fixed post length of 30 seconds, where the post score is relatively small, Analysis Of Variance (ANOVA) yielded a significant effect of friend count ($F_{7.53}$ = 25.07, $p < 0.01$), no significant effect of post score ($F_{7.53}$ = 0.22, p = 0.8) and no significant interaction effect on friend count × post score ($F_{5.12}$ = 0.23, p = 0.9).

Among these simulations with almost no reposts, higher variances in the number of reads are observed for the experiment IDs of 9-11. This is in line with the Facebook observation that "a post produced by a user with many friends has more variability in the audience size than one produced by a user with few friends" [37]. Here, part of the reason is due to the differences in the first publisher's activity parameter. For example, in simulation #9 with zero reposts, it was observed that one repetition of the experiment had a very low activity parameter of 5%; for this reason, the post remained as the most recent post of the publisher and did not slide down the wall page. Consequently, more friends had chance to read this post. In this simulation, we had the highest number of reads which in turn increased the variance of the number of reads.

Also, the time of publishing a note relative to other events at the time is another important factor to receive a high number of reads. In fact, the activity parameter and timing both represent the complexity and heterogeneity of users and their interaction within the system. We did not explicitly analyze the impact of the activity parameter in the number of reads. However we did observe that for a certain post length in cases where the post score is low, the friend count parameter has the greatest impact and the publisher's activity is the second dominant parameter in determining the audience size. The reason is that, if the post is not interesting enough to be shared by others, it would only be read by the immediate friends of the publisher. So in order to increase the number of reads in this case, a higher friend count would help. Secondly, a lower activity level by the publisher keeps the post recent and top on their wall page. This in turn increases chances of being seen by others.

TABLE III
RESULTS IN FIRST SET OF SIMULATIONS AFTER ONE WEEK

| ID | Post Score | Post Length | Friend Count | No. of Reads Mean | No. of Reads Median | No. of Reads Std. Dev. | No. of Reposts Mean | No. of Reposts Median | No. of Reposts Std. Dev. |
|---|---|---|---|---|---|---|---|---|---|
| 1 | 0.001 | 30 | 9 | 2.3 | 2 | 1.4 | 0 | 0 | 0 |
| 2 | 0.01 | 30 | 9 | 1.7 | 2 | 0.7 | 0 | 0 | 0 |
| 3 | 0.1 | 30 | 9 | 2.0 | 1 | 1.2 | 0.2 | 0 | 0.4 |
| 4 | 0.001 | 30 | 139 | 13.0 | 10 | 10.3 | 0 | 0 | 0 |
| 5 | 0.01 | 30 | 139 | 10.5 | 8 | 7.7 | 0 | 0 | 0 |
| 6 | 0.1 | 30 | 139 | 17.1 | 13 | 16.5 | 1.30 | 1 | 1.9 |
| 7 | 0.3 | 30 | 139 | 33.0 | 25 | 28.8 | 6.2 | 5 | 6.5 |
| 8 | 0.5 | 30 | 139 | 101.1 | 13 | 341.7 | 32.1 | 4 | 110.2 |
| 9 | 0.001 | 30 | 530 | 45.8 | 28 | 46.6 | 0 | 0 | 0 |
| 10 | 0.01 | 30 | 530 | 57.0 | 41 | 45.9 | 0.3 | 0 | 0.5 |
| 11 | 0.1 | 30 | 530 | 55.9 | 42 | 50.6 | 4.2 | 2 | 4.8 |
| 12 | 0.3 | 30 | 530 | 73.3 | 50 | 68.0 | 14.0 | 8 | 13.6 |
| 13 | 0.5 | 30 | 530 | 2075.5 | 1588 | 2109.5 | 702.1 | 532 | 724.1 |
| 14 | 0.1 | 60 | 530 | 30.4 | 20 | 24.1 | 2.3 | 2 | 2.2 |
| 15 | 0.3 | 60 | 530 | 33.4 | 24 | 25.9 | 6.1 | 6 | 4.6 |
| 16 | 0.5 | 60 | 530 | 41.6 | 23 | 52.7 | 13.8 | 7 | 18.5 |
| 17 | 0.1 | 90 | 530 | 12.3 | 10 | 8.2 | 0.6 | 0 | 1.1 |
| 18 | 0.3 | 90 | 530 | 12.8 | 10 | 11.7 | 2.6 | 2 | 3.1 |
| 19 | 0.5 | 90 | 530 | 14.1 | 11 | 12.5 | 3.7 | 2 | 3.8 |

The initial implication is that if one cannot make an attractive post with high interest, at minimum, one needs to have it posted by a user with large number of friends in order to reach its maximum audience.

Of further interest is knowing more about the importance of a post score versus a publisher's friend count. Assuming the note is interesting, the question is whether one should focus on finding a hub user with many friends to post it or should one improve the quality of note as much as possible. Consider the simulation IDs 6-8 and 11-13, all of which have the same post length but relatively high scores published by users with different friend counts. Among these experiments, ANOVA yielded a major effect of post score ($F_{4.8} = 19.04$, $p < 0.01$), a bit weaker but still strong effect of friend count ($F_{6.86} = 18.43$, $p < 0.01$) and major interaction effect on post score $\times$ friend count ($F_{4.8} = 16.36$, $p < 0.01$). We can conclude that in cases where the post interest and the friend count are large enough, the former parameter is more influential than the latter. Keep in mind that friend count is still important, and one needs to consider the combined effect of both together. However, if one is able to find a user with an acceptable number of friends, it is recommended to focus more on post content rather that necessarily finding a hub user with many friends. This phenomenon implies that having a good seeder may help reach/saturate a local cluster of the network faster, but ultimately a higher post score is needed to reach further regions of the network.

The next question is the trade-off between post quality and post length when we have a well-connected user with relatively high number of friends to publish the desired post. To compare the impact of post score versus post length, for simulation IDs 11-19, the number of reads is shown as a heat-map in Fig. 5. According to the figure, post length dominates post score within all the range of post scores and lengths simulated. This means in order to reach a maximum audience, keeping the message brief is more significant rather than making an impressive but lengthy one. For example, by comparing the simulations #11 and #16, both of which have similar number of reads, it is observed that for a long note, more reposts (and consequently more time) is required to reach a similar audience size of a short note with a lesser score. Statistically speaking, for the simulation IDs 11-19, ANOVA yielded a strong effect of post score ($F_{4.73} = 18.33$, $p < 0.01$), a bit stronger effect of post length ($F_{4.73} = 20.35$, $p < 0.01$) and a significant interaction effect on post score $\times$ post length ($F_{3.43} = 18.03$, $p < 0.01$). Therefore both post length and post score are important properties, and their combined effect has to be considered when making a post; yet, depending on the situation, the post length can be considered of more importance as if the length extends over a certain threshold it severely affects the post reachability no matter what the post score is. The reason is that users generally do not spend much time to judge a post. For example, a post might be very amusing, but as it is lengthy a typical user never spend sufficient time to recognize true score of the post.

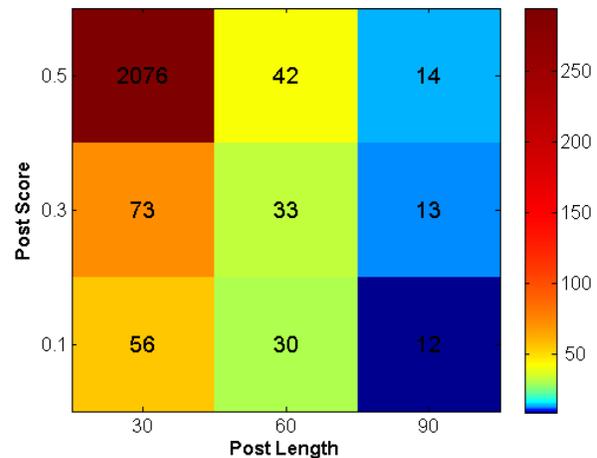

Fig. 5. Impact of Post Score versus Post Length on number of reads (medium to long notes)

TABLE IV
RESULTS IN SECOND SET OF EXPEREMINTS AFTER ONE DAY

| ID | Post Score | Post Length | Friend Count | No. of Reads | | | No. of Reposts | | |
|---|---|---|---|---|---|---|---|---|---|
| | | | | Mean | Median | Std. Dev. | Mean | Median | Std. Dev. |
| 20 | 0.001 | 30 | 319 | 33.8 | 28 | 25.4 | 0 | 0 | 0 |
| 21 | 0.1 | 30 | 319 | 35.2 | 19 | 32.9 | 2.6 | 2 | 2.5 |
| 22 | 0.25 | 30 | 319 | 48.4 | 41 | 26.4 | 7.6 | 7 | 3.9 |
| 23 | 0.5 | 30 | 319 | 1857.2 | 58 | 2465.4 | 635.8 | 17 | 852.7 |
| 24 | 0.75 | 30 | 319 | 7651.8 | 6670 | 8085.1 | 3815.3 | 3324 | 4035.3 |
| 25 | 1 | 30 | 319 | 23559 | 23468 | 507.6 | 15325.6 | 15213 | 340.8 |
| 26 | 0.001 | 10 | 319 | 46.7 | 46 | 12.3 | 0 | 0 | 0 |
| 27 | 0.1 | 10 | 319 | 92.2 | 101 | 55.2 | 5.1 | 4 | 4.7 |
| 29 | 0.25 | 10 | 319 | 120.7 | 91 | 95.7 | 18.1 | 10 | 18.9 |
| 30 | 0.5 | 10 | 319 | 12449.9 | 16964 | 8,608 | 4077.9 | 5565 | 2,825.3 |
| 31 | 0.75 | 10 | 319 | 32473.5 | 32297 | 1,007.3 | 15375 | 15256 | 457 |
| 32 | 1 | 10 | 319 | 41165.8 | 41036 | 1071.6 | 25424.1 | 25330 | 587.5 |

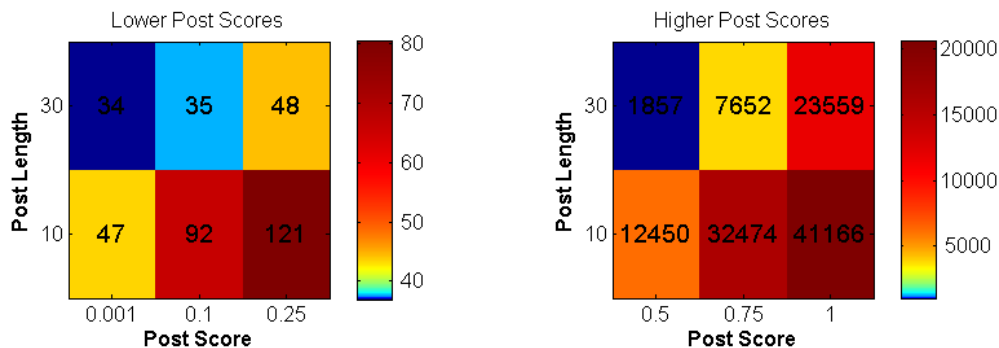

Fig. 6. Impact of Post Score versus Post Length on number of reads (short notes)

*B. Study 2: Importance of Score versus Length in Short Posts*

After inferring that a post should not be too long if it is to propagate, in the second study, 12 more scenarios were simulated to test a broad scale of different post scores versus relatively short post lengths. In this set of simulations, the user who publishes the post for first time is a new user with a different position in the network and a subsequently changed local network structure. This ensures our results are not specific only to some local region of the network structure.

This time, a total of 12 unique simulations were set up and each simulation was repeated 10 times, for a total of 120 runs for two simulated days each. In each simulation, the post spread began after a 24-hour initial warm-up phase, and then one day later the outputs were saved and simulation ended. In these experiments, smaller post lengths of 10 and 30 seconds with higher chances of sharing were tested. The post score changed from 0.001 to 1 to study a broader scale of scores. The results of our second set of experiments are shown in Table IV.

According to Table IV, as a post score increases from 0.25 to 0.5, a significant change in the number of reads is observed. Obviously, post score should likely have the greatest impact in general. For a more detailed comparison, the simulation is split into two subsections of low and high post scores. One heat-map for each part is shown in Fig. 6. From the heat-map on the left corresponding to low score posts, it can be seen that post length dominates the post score. On average, all the scenarios of shorter (i.e., 10 seconds) posts reach comparable size or larger audiences than longer (i.e., 30 seconds) posts with any scores, as long as post score is not very high (less than 25%). This verifies the previous result with a different seeder and time to collect the result. However, for the heat-map on right, this inference does not hold true any longer. For higher score posts, shorter posts do not necessarily reach more users, and it depends on both post score and post length together. For example, simulation #25 with a post length of 30 seconds and score of 100% has reached a larger audience than simulation #30 with a shorter post length of 10 seconds but lower score of 50%. Yet simulation #31 with a short post length of 10 seconds and score of 75% has larger audience than simulation #30. From an ANOVA perspective, a significant effect of post length ($F_{6.88} = 190.99$, $p < 0.01$), a more significant effect of post score ($F_{3.19} = 293.77$, $p < 0.01$) and a significant interaction effect on post score × post length ($F_{3.37} = 46$, $p < 0.01$) is observed. Therefore, generally both post score and post length are very important in the success of a post propagating. However, the relationship with these two parameters and audience size is not linear.

Assuming it would be desirable to explain impacts of each factor in the number of reads with a linear model, Table V

TABLE V
COEFFICIENTS OF LINEAR MODEL OF NUMBER OF READS FITTED TO SECOND STUDY

| Model input | Coefficients | Std. Error | Standardized Coefficients | t Stat | P-value |
|---|---|---|---|---|---|
| Intercept (constant) | 3552.302 | 1981.203 | | 1.793 | 0.076 |
| Post Score* | 33077.74 | 1788.742 | 0.820 | 18.492 | 2.2E-36 |
| Post Length* | -441.586 | 63.51394 | -0.308 | -6.953 | 2.25E-10 |
| Activity | 1923.83 | 2441.076 | 0.035 | 0.788 | 0.432 |

*Significant at an alpha level of 0.01

TABLE VI
COEFFICIENTS OF TRANSFORMED INPUTS OF LINEAR MODEL OF NUMBER OF READS FITTED TO SECOND STUDY

| Model input | Coefficients | Std. Error | Standardized Coefficients | t Stat | P-value |
|---|---|---|---|---|---|
| Intercept (constant) | -9029.287 | 1353.794 | | -6.670 | ≤0.001 |
| (Post Score)$^2$ | 3.396 | 0.158 | 0.849 | 21.443 | ≤0.001 |
| 1 / (Post Length) | 132908.250 | 17021.300 | 0.309 | 7.808 | ≤0.001 |

TABLE VII
COEFFICIENTS OF GENERAL MODEL OF NUMBER OF READS FOR THOSE POSTS FOR WHICH NUMBER OF REPOSTS ≥ 1

| Model input | Coefficients | Std. Error | Standardized Coefficients | t Stat | P-value |
|---|---|---|---|---|---|
| Intercept (constant) | 27952.530 | 69.463 | | 402.410 | |
| √ (Friend Count) | 149.849 | 1.636 | .094 | 91.583 | ≤0.001 |
| Log (Post Length) | -21236.924 | 42.333 | -.510 | -501.659 | ≤0.001 |
| (Post Score)$^2$ | 1.905 | .004 | .553 | 542.076 | ≤0.001 |

shows the result of fitting a linear model to our second study. The linear regression model has an adjusted $R^2$ of 76%. However, expectedly, the plot of residuals does not suggest a linear model as being a suitable model for these scenarios. The coefficients statistics confirm our previous inference using ANOVA and heat-maps. According to Table V, considering a α significance level of 1%, the activity parameter does not have any significant impact on the number of reads with a p-value of 0.4. Both post score and post length have significant impacts. Post score has higher influence within all the simulations studied in the second set of simulations together.

In order to obtain a better model of number of reads, each of the following (five) functions was applied to our model inputs: Inverse, Logarithm, Square root, Square and Cube. So instead of three inputs of Post Score, Post Length and Activity, there are now 3×6 = 18 inputs to choose from. The best linear model based on all these inputs was found to be a linear model of post score squared and inverse of post length. The new model has a better adjusted $R^2$ of 81% with the coefficients shown in Table VI.

Properties of over one million posts were saved after two days from the starting point of simulations in different runs for another linear regression. The basic inputs were post score, post length and friend count. However, once again the five functions of Inverse, Logarithm, Square root, Square and Cube were applied to each of the basic inputs to find the best combination of inputs for the regression model. The best model was found to have an adjusted $R^2$ of only 32%. Subsequently, posts with zero repost were excluded from our input data to have more coherent input data, which resulted in almost half a million posts remaining. With the objective to have only one function of each of the basic inputs as an input to our linear model, the best fitted model was found to be a linear model of post score squared, logarithm of post length and square root of friend count. This model has an adjusted $R^2$ of 54%. The coefficients of the model are shown in Table VII. The scatterplot of standardized residuals versus standardized predicted value is shown in Fig. 7. Ideally it would be desirable to have a uniform scattering of points around the zero reference line; but as the agent-based model is highly non-linear, a better linear model could not be fit to its outcome.

All the inputs have coefficients significantly different from zero. Additionally, by looking at the standardized coefficients, it is clear that generally both 'logarithm of post length' and 'post score squared' have significant impacts on number read. Yet post score squared is a stronger predictor.

According to the second study, one needs to avoid lengthy posts, and the post still needs to have some minimum score. There are certain thresholds for post lengths and post scores that post properties should lie within. Initially, it is better to focus on post score to gain some interest and chances of sharing, then one should try to shorten the post length considering the limited time of users. Lastly, when the message is brief enough, fine-tuning the post score can achieve better results than slightly reducing of post length.

There is an uncommon difference between the median and average number of reads in simulation #23. The low median number of reads can suggest that a typical user may not read a long post. However if, with the help of good timing, users read a long (but interesting) post, the post would propagate very well through the network increasing the total number of reads. Therefore in simulation #23 when the timing is matched, we observe a high number of reads increasing the average number of reads, and when timing does not cooperate well, users show little interest in the post.

*C. Study 3: Comparison of two Seeding Strategies*

In the third and last study, the objective was to gain insight by comparing a classical seeding strategy of a small number of hub users with many friends versus seeding a large of number users with few friends. First, a special post was published by four randomly chosen users in the network, each of which has exactly 50 friends. Then, the simulation was repeated for two randomly chosen users with 100 friends each. The details are described below.

All four users had a fixed activity level of 50%. At a certain time after the initial warm-up phase, they all shared a unique short post with 50% score and 10 seconds length. Then 24 simulation hours after the spread, the numbers of reads were collected and saved. The simulation was repeated 10 times. Then the whole simulation was repeated for two more sets of four random users with exactly 50 friends, for a total of 30 runs for the 4×50-friend case. The same simulation was carried out for three different sets of two users with exactly 100 users each, for a total of 30 runs for the 2×100-friend case. The average number read for the 4×50-friend case and the 2×100-friend case is 13300.7 and 9540.1, respectively. The statistics are shown in Table VIII.

The average results suggest that a mass of small seeders may broadcast a certain post better than a few hub users. However, assuming the more general case of not-equal variances for these two cases, one cannot reject the null hypothesis of having equal means in these two cases. Technically, the null cannot be rejected by t-test with a t stat of 1.679 and the degrees of freedom of 57.426. This is equivalent to a non-significant p-value of 0.099. In other words, the difference between the two cases is not statistically significantly different.

To see the trend of message propagation, Fig. 8 plots the number of reads and number of reposts versus time for a run of simulation #13 on a log-log scale. All other simulations where the post is shared by some users have a similar trend. The only time required for the outbreak to propagation is the seconds required to read the post. Unlike other diffusion patterns such as product adoption or the spreads of infectious disease, this trend is not s-shaped for Facebook posts. In other words, there is no classical tipping point or epidemic threshold for post propagation after which we could expect an outbreak in the number read. The reasons are discussed in section IV. This trend may sound surprising at first, yet it is consistent with real observations of Facebook. According to a select group of brand posts data of Facebook in November 2012, each post, on average, reached half of their target audience within 30 minutes after publication [43]. Also a quite similar trend is reported for the number of retweets in Twitter [44].

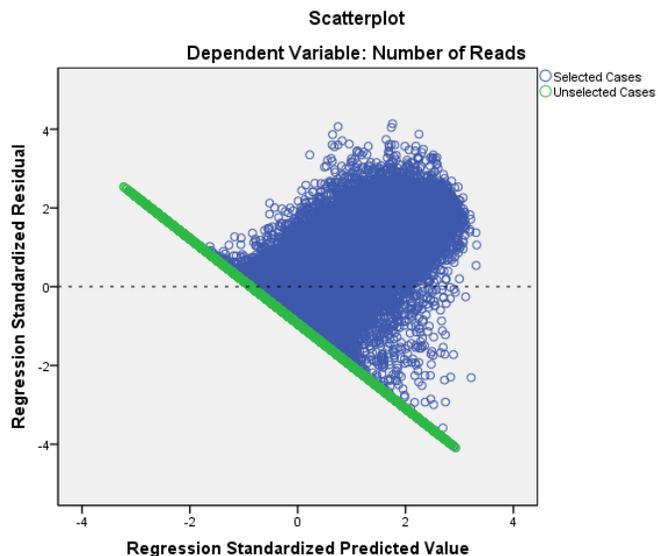

Fig. 7. Scatterplot of standardized residuals of linear model versus standardized predicted value of number of reads

TABLE VIII
STATISTICS OF NUMBER OF READS IN THIRD STUDY

| Case | Mean | Std. Deviation | Std. Error Mean |
|---|---|---|---|
| 4×50-friend | 13300.73 | 8228.25 | 1502.27 |
| 2×100-friend | 9540.10 | 9096.65 | 1660.81 |

Another surprising result is that the median number of reposts is zero. This means it is quite common that posts by a typical user on Facebook do not attain even a single repost. We confirm that there exist posts with high number of reposts in both Facebook and our agent-based model. These highly shared contents are mostly (high scored) posts published by popular pages, which can be thought of as hub users in our agent-based model. Yet many of the posts, especially the ones submitted by typical users receive few or very limited reposts. Most of the published results about social networks are generally focused on successful posts and their properties, and as such, statistics regarding the failed ones submitted by random usual users could not be found. The total distribution of number of reposts is shown in Fig. 9. This long-tailed distribution indicates only a few posts gain a huge number of reads. The distribution is also consistent on the surface with that of retweets (popularity) for twitter reported by [44]. Section IV discusses this consistency.

Initially, we expected that having a score of 10% would be enough for a post to be broadcast and seen by everybody. Firstly, in this case, the average probability of reposting is 1/20. Secondly, when the post is published for the first time, only the immediate friends will see the post. The immediate friends should be online in a certain time period to catch the post. Therefore, if the friend count is not sufficient, the post will never have any chances to spread further. In other words, if the network were fully connected so that users observe the whole population, the reposting probability of 1/20 might have been adequate, however this is not the case in this agent-based model.

## IV. RELATED WORK AND DISCUSSION

Diffusion is a core process in many areas such as energy flows in physics, disease spread in biology, behavioral contagion in sociology and product adoption in economics. Several researchers have studied aspects of this phenomenon in human systems. In human systems, a general way of social contagion is through the word-of-mouth (WOM) mechanism which is quite similar to spread of an infectious disease within a population. Below, the current simulation results are discussed in relation to other common diffusion processes of product adoption (marketing), disease spread (epidemiology), and electronic networks.

Within product adoption research, Moldovan *et al.* studied the importance of WOM to new product success [1]. In their research, a product score is divided into two dimensions of originility and usefulness of a product, and the number of online reviews is used as a proxy for the amount of WOM. Reference [5], using an ABM, studies the actual value of a seeding program for WOM in terms of market expansion and purchase acceleration for a certain product. Reference [36] attempted to predict users' adoption of a given product on the Digg social news website. They studied the effect of network-level dynamics and changes in individual preferences for different topics, and proposed a viral marketing strategy which was tested with an agent-based model. Reference [45] studies social commerce where individual sellers are connected to each other through an online social network. Using time series analysis at the marketplace level and Bayesian statistical analysis at the shop level, they explore whether connecting the sellers to each other increase the sale, and how the position of a seller within the networks influences their value. Reference [46] studies impact of different connection patterns among individuals on the diffusion process in a European online social network. Using a hazard-rate model, they investigate characteristics and local structure of potential adopters and their neighbours.

Tirunillai and Tellis also confirm the importance of user-generated content as a type of WOM and in particular study the impact of online product reviews and ratings on stock market performance using multivariate time-series models [7]. Similarly, [3] finds that consumer generated online product ratings has a direct relation with the product sales; although the previously submitted ratings affect the future ratings. Reference [47] also investigates the evolution of online reviews of books over time and sequence. Weblogs as a part of the larger set of online social media can also influence purchase/adoption of a product. An analytical model for a blogger is made and studied in [48]. In particular, they explain why blogs may link to rivals, and what the relative benefits and costs of such linking are.

Goldenberg *et al.* study the role of hubs in the diffusion process of products over the Cyworld social networking site in

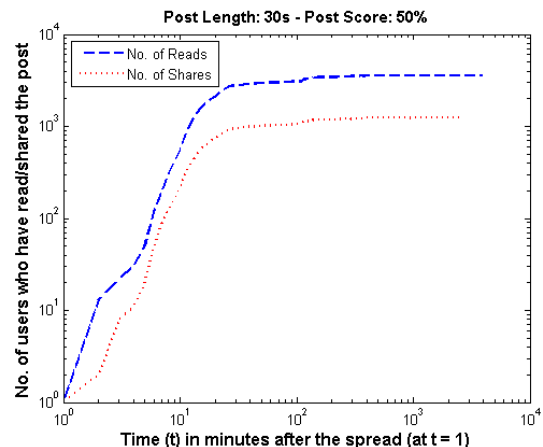
Fig. 8. Trend of message propagation

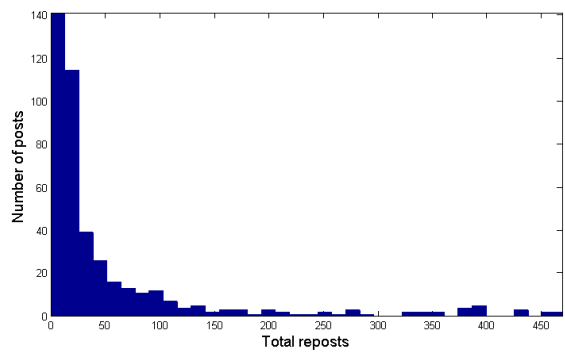
Fig. 9. Reposts distribution

Korea [49]. They define two types of innovator hubs who need little exposure to adopt a product and follower (social) hubs who are well-connected. Using an ABM, they analyzed the impact of each type on the market eventual size and the speed of adoption.

Cellular automata modeling and aggregate level modeling together are used in [50] to study growth rate of a new product. By separating network externalities effects such as mass media and advertising from internal interactions (i.e., word of mouth through the network), they found a chilling effect on growth rate of a product. This is a "wait-and-see" state for a product when potential consumers wait for others to adopt a certain product and then decide whether to purchase or not. This partially explains why product adoption has an s-shaped growth rate. Thus growth of products can follow a two-stage process which includes a slow start due to the chilling effect and then a fast growth because of the bandwagon effect. However, a similar concept for post-sharing does not exist. One does not need to see if others have shared a post to decide whether one should or not; basically because one is not purchasing a post with real currency.

In the context of contagious disease, [51] studies SIR models of epidemic disease spread in networks. Rahmandad and Sterman compare AB and DE models of epidemic spread on different networks [52]. There are some similarities and differences between the spread of an electronic post and an infectious disease. The propagation is similar in the way in which it cascades through the network. Unlike posts and similar to products, disease has an s-shaped spread rate (for basic review, see [53]). In a disease epidemic, the rate of people entering from a susceptible into the infected state depends on the number of people in other states (such as the infective state). As people become more and more infected, chance of transmission of disease is increasing until the epidemic threshold is passed and the disease rapidly spreads throughout the population. In our agent-based model, it is assumed that a post score, corresponding to chances of reposting, is constant. In epidemiology, a quicker recovery rate makes a smaller epidemic; it takes a longer time to happen as people recover faster and the number of infected in the beginning of the process is not enough to form an outbreak. In our agent-based model, users need to read a post completely (i.e., become infected) then may spread it. With a short post, a shorter time is expected for the outbreak to occur, and it has a higher peak, as the time needed to read the post is smaller and more people are likely to read it. During the infection period of the disease epidemic case, the chance of transmission of the disease is constantly present; so the epidemics with a slower recovery (removal) rate (i.e., a long infected period) that slowly kill people are more dangerous to the populations having more deaths at the end (for discussion see [53] pages 21-22). In social networks, users generally share a post only once and their friends usually read the post once and decide whether to share or not. As the assumption is people do not share a post while reading it, a longer post (unlike a longer infected period of a disease) does not provide a better opportunity to spread the post. Lastly, the ultimate objective in disease modelling is mitigating against the spread of the disease such as targeted vaccinations as a means of achieving herd immunity; but in social networks, more spread and penetration are desired.

Diffusion of applications in Facebook is studied in [54] using a customized commercial application about the movie industry to track user behaviour. Using hazard modeling, they test effectiveness of passive-broadcast messaging versus active-personalized messaging. In another study on Facebook, susceptibility of various type of users (e.g. young, women, married etc.) to influence their adoption decisions is measured [55]. Application adoption and social influence in Facebook is also studied in [56] using fluctuation scaling (FS) method. They track popularity of a set of applications among all users in their dataset collected in 2007. Their observation is limited to the adoption of an application, and not necessarily the usage. Similarly, sharing a post needs an increased level of engagement, rather than simply reading it.

Reference [57] analyzed the prevalence data reported by a computer virus for a time window of 50 months. They found the absence of an epidemic threshold for virus spread on scale-free networks due to an infinite connectivity effect phenomenon in large scale-free networks. This effect happens because of a heterogeneous rising and falling in the number of links of nodes in the scale-free graph. This factor is also applicable to the current agent-based model as the synthesized graph shares this feature of scale-free networks.

There are various and sometimes conflicting recommendations regarding choosing the optimal set of users to publish a message (see [58] for summary of previous research). Reference [58] analyzes four different seeding strategies for both messages and products using two small-scale field experiments and one real-life viral marketing campaign. Their four seeding strategies are: 1) seeding well-connected hub users; 2) seeding low-degree members, called the fringes; 3) seeding high-betweenness users, called the bridges that connect different sub-networks; and 4) random seeding. They find high-degree and high-centrality strategies are preferable in general. In contrast, [59] finds large cascades are driven by critical mass of individuals, but not necessarily influentials (hubs). In the current third simulation, some evidence in favor of this strategy of mass of individuals also emerged. However, the test was limited and the results were not significantly different from seeding a few hub users. In summary, it is safe to conclude each strategy may work well under certain conditions.

Furthermore, in the current agent-based model, by investigating properties of the most successful posts, it was found that as long as a post is not qualified enough to be reposted by typical users, it does not have any chance to diffuse across the whole network. In other words, relying only on reposts by active users (with high tendency to share) does not guarantee message propagation. Because wall pages of active users who repost many notes tend to be crowded with different notes, the note would only have a minimal chance to be read by someone else. Nevertheless, if the note is reposted by a selective hub user with many connections but low activity to post other notes, a significant growth in the number of reads and reposts will be achieved. As the post remains as the most recent post of the hub, it gains maximum exposure to the hub's friends, although one still needs the post to be highly scored to be ultimately reposted by a critical mass of users. Therefore, to achieve a maximum audience, it is recommended to produce the desired post and begin its publishing by different hubs who have many connections but (who) are quite discerning about their willingness to share other notes.

Reference [60] analyzes the impact of the degree distribution of a network on adoption process, using an agent-based model. They discovered that while most researchers simply assume the adoption process propagates over the entire visible (overt) network, the actual active subset of the network over which the propagation occurs may have quite different properties from the entire network. They provided evidence that the degree distribution of the active network is generally different from that of the entire network. The degree distribution has a significant effect on contagion properties of nodes within the network.

In an inspiring paper by Lerman and Ghosh, the active network of users on Digg and Twitter is extracted, and then it is studied to see how the structure of the (active) network affects the dynamics of information flow on each network [44]. The general mechanism of the spread of information in both sites are similar to Facebook, where users watch their friends' activity, and they may share/tweet/vote for a post to make the post visible to their own friends/fans/followers. The precise underlying details vary from case to case though. As mentioned in section III, the evolution of number of tweets received by each post in Twitter reported in [44] is utterly similar to that of reposts in the current work as displayed in Fig. 8; (successful) posts display a burst of growth at the beginning, and the growth saturates after a while. The point at which growth saturates is the cascade size indicating how far and wide the post has penetrated though the network. Reference [44] reports that Digg's (active) social network of their dataset has a larger clustering coefficient than Twitter's, meaning that its network is denser. As such, initially posts spread easier in the more highly interconnected Digg's network, but they eventually spread more distance in Twitter's less densely connected network. The distribution reported for the number of retweets (or reposts) in [44] is somewhat analogous to the distribution of number of reposts in the current agent-based model displayed in Fig. 9. Retweets distribution in Twitter has a small number of posts (or tweets) with almost zero retweets, followed by an exponential peak and then a gradual decrease of retweets to zero again, creating a long tail. This means a majority of posts gain a few (but non-zero) retweets in Twitter and very few posts exist with many retweets. Fig. 9 states the same fact finding, except that it claims that the majority of posts in Facebook have almost zero reposts. The reason is that the Twitter dataset used in [44] does not include any non-popular tweets (with low score). They collected only frequently-retweeted posts on Twitter using the Tweetmeme.com site at the time. In addition, tweets can only have a limited number of characters. As such, they all can be considered as short-length posts as in the current agent-based model, with higher chances of being reposted. However the distribution reported in Fig. 9 corresponds to all the posts with various length and score simulated within the agent-based model.

Reference [61] employs an agent-based model to simulate reposting behaviour of users in the Twitter social network. They explicitly model the competition for humans' limited attention among different posts (memes), and how it affects memes' popularity. They developed a memory mechanism to reflect users' past behaviour to what they would share in the future, as users are likely to share posts similar to what they shared before. Similar to the result we presented in this paper, long-tailed distributions for memes' popularity and lifetime are reported. They found extremely heterogeneous behaviour; a few memes are extremely successful while most of them die out quickly. Reference [62] proposes a coupled hidden Markov model to capture neighbour users' influence on users' posting activity. Their model is also tested on Twitter.

An interesting study on information cascades on Digg by Steeg *et al.* revealed that a high level of clustering structure of the Digg network limits the overall growth of cascade [40]. In highly clustered networks, people are usually exposed to a certain post multiple times through multiple friends, which in turn lowers the epidemic threshold, speeding the spread up initially. However, the surprising effect reported by [40] is that repeated exposure to the same post does not encourage people to repost it. This is a fundamental difference between the spread of information and classical spread of products or diseases. Reference [40] shows this effect drastically limits the overall cascade/epidemic size. As an example, while many posts with a fast starting spread exist in their dataset, only one post about Michael Jackson's death reached a significant fraction of 5% of active Digg users. They observed that the effective number of people who have not been exposed to a post is gradually decreasing. In addition, other effects such as decay of visibility and novelty could be other reasons why the epidemic stops [40]. In one more study on Twitter, rapid decay of visibility combined with the limited attention of users are determined to be the primary reasons for preventing the growth of propagation of online information [15].

The way the current model network described in this paper is synthesized does not produce a high (and desired) level of clustering coefficient. This means that the number of reads for successful posts may have been overestimated. That being said, most successful posts in the current simulations end up reaching around 30% of the entire network. The percentage of viral posts is less than 2.5% of all the posts generated within the model. Moreover, it was discovered that all these posts had extremely high scores of mostly over 90% and short lengths of mostly 10-15 seconds published by users with various friend counts. Precisely speaking, these viral posts have scores ranged from 70% to 100% with a mean and median of 89% and 91%, respectively. The post lengths are varied from 10 to 25 seconds with a mean and median of 15 and 14.5 seconds, respectively. It implies that viral messages can be published by a user with low number of friends, but certainly various users including hubs would have to repost it during the spread. More importantly, the post must be brief while extremely highly qualified. We used a singular numerical value to represent a post score, but a post can have different aspects to be engaging to different kinds of people. There are various categories of posts: promotional offers/deals, advice and tips, warnings, amusing video clips, amazing pictures, personal news, motivating speeches, campaigns recalls etc. Thus, a score of 90% in our simulations represents a high quality post in a variety of features. But in reality, it is nearly impossible to make such a universally fascinating post. Certainly, viral posts exhibit a variety of features. For example, death news of a famous celebrity has a wide range of viral features making it likely to go viral. Such a post has information, novelty, may contain stimulating quotes or represent a group of people's mourning or respect.

## V. CONCLUSION AND LIMITATIONS

There has been a great deal of work on dissemination of information in online social networks especially on Twitter and Digg. Studies on application adoption in Facebook also exist. This paper presents a large scale stochastic agent-based approach for modeling wall post propagation within the Facebook network. Network and other input parameters have been drawn from and tuned to published sources of Facebook statistics. Other studies on the realm of social media diffusion have confirmed the importance of various factors including underlying network structure, local network structure of following spreaders, influence degree and activity level of each spreader, type and novelty of a post and people's response to repeated exposure. At the current stage, analyzing all these factors in a single study would be too complex or too generic. The agent-based platform created here is potentially capable of testing all factors by some code modifications. In some cases, one would have to insert a few more parameters into the system, such as a social influence parameter for each person for example.

In this paper, various scenarios have been explored to investigate the impact of post length, post score, and the post publisher's friend count on post diffusion. It is observed that posts with small scores hardly ever spread farther than the immediate friends of the publisher, meaning that higher friend count for the publisher has a stronger impact on audience size than higher post score does. However, beyond a certain level of message quality, the post score has a larger influence on message propagation through the network than publisher's friend count. This intuitively means the content of a message is more important than who has delivered it. In cases with relatively medium post interests and medium friend counts where there definitely exists some reposting, the post length is the most influential parameter followed by the post score and the friend count as the second and third influential factors, respectively. This implies that creating a long post makes it boring and significantly affects its chances of getting shared by others. Whereas for relatively short posts, increasing the quality of post contributes more to the audience size than cutting the post length any shorter, which does not necessarily boost the growth of number of reads. The intuition behind this result is that people spend a minimum amount of time on each post, and once a post length is below that minimum length, there would be no need to make the post any shorter. Keeping these hints in mind could help marketers to find a balance between length and content of post making their ideal post advert for example.

Adjustments are needed to study other online social networks using this agent-based model. For example, reconstructing the network, limiting post lengths and the way users look through the posts from friends have to be changed in order to study the Twitter social network. However, the results reported here may not be limited only to the Facebook online social network. Less intuitive findings about the dynamic of post spread mechanism such as lack of epidemic threshold in the propagation of information have been previously confirmed in other online social networks. Along with other work on seeding strategies, it is observed that both hub-seeding and having a large number of individual seeders could result in having a viral post reaching an epidemic portion of population of at least a sub-region of the network graph. The simulations performed to compare these two strategies were limited to conclude a general statement though.

Achieving an 'epidemic reach' of the entire Facebook network is nearly impossible. People are online at different times and it is vital for a post to catch their attention when they are online. Assuming the timing can be handled through the interface design of online social networking websites, a viral post still needs to be highly scored in a variety of features. Each person is likely to become engaged in a certain category of posts. Modeling a post score based on a single scalar value may be an over simplistic assumption. A much more realistic way to define a post score is using a vector where each component describes post content in a different perspective. Then users' activity parameters also need to be vectorized to capture the heterogeneity of population in different directions. Add to this the fact that certain categories of contents (e.g. politics, fashions, and sports) are being shared more on certain social networking sites (e.g. Twitter, Pinterest, and Facebook) [62]. The limitation in the current paper is that an active user is likely to share any type/category of posts, and also a high-scored post has all the features of all sorts of appealing posts which cannot be true in practice.

One more realistic extension of current research would be the insertion of dynamic scores for post. There are different ways and reasons to change a post's score during the spread. A rational reason is that after users spend a few seconds on a post, they get a better idea of how interesting the post is. So they may decide to continue reading/watching the post or disregard it. This is especially true for video posts. Once the dynamic scoring feature is added into the model, one can let users share a post even before they read it completely.

There are other limitations with the current study which bring opportunities for further extensions of this research. Sensitivity analyses for input parameters such as average post length or login rate were limited only to the primary stages of making the agent-based model. The parameters and distributions were tuned on the basis of a smaller network and Facebook's known statistics. The parameters were fixed once it was observed that the main agent-based model of the larger network had rational functionality in line with the conceptual model. Testing sensitivity of all the outputs to all input parameters on such a large set of data would require a lot of time and effort. Such reports could offer insights on what to expect if people begin spending twice as much time on Facebook for example, or investigate robustness of results across a range of parameters and distributions. In addition, the exact correlations, threshold values and interactive effects of simulation inputs (post length, post score and publisher's friend count) still need to be determined through more simulations.

One crucial direction for future work is a more comprehensive way of modeling the Facebook news feed ranking algorithm, which used to be called the EdgeRank algorithm. Our current model of EdgeRank emphasizes Recency of posts too highly, meaning that if a post is recently published, it has the highest chance to be seen by users. Facebook tries to identify those who have most interaction with a user. Currently the algorithm is considering numerous factors (including recency) to decide which posts to show for each person in their home news page. For example, if one likes or comments on posts by a person/group, chances of receiving more posts from those persons/groups will be increased. So modeling these two Facebook features could be an essential feature to complement the current agent-based model.

APPENDIX

Let $U = AB$, where $A = U[0,2]$ and $B = N(\mu, \sigma^2)$ are independent random variables (rvs). Using the following formula for the Probability Density Function (PDF) of product of two independent rvs,

$$f_{U=AB}(u) = \int_{-\infty}^{\infty} \frac{1}{|x|} f_A(x) f_B\left(\frac{u}{x}\right) dx \tag{1}$$

we can find that in our case

$$f_U(u) \begin{cases} \frac{1}{\sqrt{2\pi}} \int_{\frac{u-\mu}{\sigma}}^{\infty} \frac{1}{x*\sigma+\mu} e^{-\frac{x^2}{2}} dx & ; u > 0 \\ \frac{-1}{\sqrt{2\pi}} \int_{-\infty}^{\frac{u-\mu}{\sigma}} \frac{1}{x*\sigma+\mu} e^{-\frac{x^2}{2}} dx & ; u < 0 \end{cases} \tag{2}$$

This density is given in terms of the integral that cannot be evaluated explicitly but can be approximated numerically.

ACKNOWLEDGMENT

The authors would like to thank Professor Miroslaw Pawlak for his input and guidance for the mathematical analysis.